
\normalbaselineskip=20pt
\baselineskip=20pt
\magnification=1200
\overfullrule=0pt
\hsize 16.0true cm
\vsize 22.0true cm
\def\ut#1{$\underline{\smash{\vphantom{y}\hbox{#1}}}$}
\def\lam{{ \Lambda^2 }}
\def\Psihat{{ \widehat \Psi }}
\def\phat{{\widehat p}}
\def\Mhat{{\widehat M}}
\def\Uhat{{ \widehat{\cal U} }}
\def\U{{\widehat U}}
\def\lambar{{\overline{\lambda}}}
\def\bop2{{ {{ {\widehat b}^2}\over {b_H^2}} }}
\def\B {{ \vec B }}
\def\ohat{{ \widehat \varphi }}
\def\so2{{ {{\sigma}\over 2} }}
\def\lsim{\mathrel{\rlap{\lower4pt\hbox{\hskip1pt$\sim$}}
    \raise1pt\hbox{$<$}}}         
\def\iout#1#2{{ \int_{-\infty}^{#1} d#2 \rho(\B,#2) }}
\centerline{{\bf DRAFT 7 august7.tex}}
\centerline{{\bf  SUM RULE DESCRIPTION OF COLOR TRANSPARENCY}}
\vskip 18pt
\centerline {L. Frankfurt$^1$}
  \vskip 12pt
  \centerline{NSCL/Cyclotron Laboratory}
  \centerline{Michigan State
  University, East Lansing, MI 48824-1321}
  \vskip 12pt
\centerline{ W.R. Greenberg, G.A. Miller}
  \vskip 12pt
  \centerline{Physics Department, FM-15}
  \centerline{University of Washington, Seattle, Washington 98195}
   \vskip 12pt
   \centerline{and  M. Strikman$^{2}$}
   \vskip 12pt
   \centerline{Department of Physics}
   \centerline{Pennsylvania State
   University, University Park, PA 16802}
\vskip 12pt
\centerline{{\bf ABSTRACT}}
The assumption that a small point-like configuration does not interact with
nucleons leads to a new set of sum rules that
are interpreted as models of the baryon-nucleon interaction.
These models are rendered semi-realistic by
requiring consistency with data for cross section
fluctuations in
proton-proton diffractive collisions.
\vskip 1.0truein
\noindent $^1$ On leave from St.Petersburg Nuclear Physics Institute,
address from Aug. 1, 1992- Jan. 31, 1993 -Physics Department, FM-15
University of Washington, Seattle, Wa.  98195
\noindent $^2$ Also at St.Petersburg Nuclear Physics Institute

\vfill \eject
\noindent  1. \ut{Introduction}

Color Transparency (CT) is the  suppression of
final (and/or initial) state
interactions  caused by the cancellation of color fields of a system of
closely separated quarks and gluons~[1,2].
For example, in an (e,e'p) reaction with good kinematics
the proton wave will be better approximated by a plane wave than by
the usual exponentially decreasing wave function.
The kinematics are ``good'' if
the  nuclear excitation energy
is known well enough
to ensure that no extra pions are created.
With these kinematics  the final state
interaction is either elastic, or a
two-body reaction.
Such reactions are topics of active experimental investigation [3-5].

 Color transparency CT occurs if: (1) a small-sized wave
packet (point-like configuration or PLC)
is formed in a high momentum transfer reaction~[1,2], (2) the
interaction between such a small
object and nucleons is suppressed
(color neutrality or color screening in perturbative QCD~[6],
the composite nature of hadrons in nonperturbative QCD~[7]),
(3) the wave packet escapes the nucleus while
still small [8].

The evidence for item (2) is well-documented, so we take it
as given. The
effects of color neutrality (screening)
in hard processes are reviewed
in Refs [7,9]. Attempts to apply  the two
gluon exchange model to soft
processes are discussed in Ref.~[10].

 The truth or falsehood of item (1) at
experimentally feasible energies
is an interesting issue  discussed in Refs. [9,11,12]. Here we take the
formation of a PLC as a starting point.

The escape process is our present concern. At infinitely  large
momentum transfer, CT is a consequence of
quantum chromodynamics QCD, see e.g.Ref.~[13].
For very large but finite energies, the cross sections, for processes we
consider,
decrease rapidly with momentum transfer. So,
the feasibility  of searching for color transparency becomes questionable,
unless sub-asymptotic momentum transfer processes can be interpreted.
For such kinematics,
the $PLC$ wave packet
undergoes time evolution which increases
its size. The time
$\tau$ to achieve characteristic hadronic size is $\tau \approx E/m\tau_0,$
where $\tau_0 \approx$~1~fm is a characteristic rest
frame time, and
$E$ is the lab energy.
The mass $m$
may be roughly the average of the masses
of the nucleon and its first excitation according to Refs.~[8,9,12,14,15].
So, for presently available energies,
$\tau\sim 5 \tau_0 \approx$ 5 fm is small enough so that
the $PLC$ expands significantly as it moves through the nucleus.
Thus the final state interaction is suppressed but does not disappear;
its effects are
studied in Refs.~[8,9,14-15]. For example,
the use of  pQCD and Bjorken scaling for deep inelastic
processes [8] allows one
to obtain the dependence of PLC-nucleon cross section on the
distance from the point of the hard interaction.
 Another procedure is to decompose the PLC as a coherent sum
over hadronic states and to study
the propagation of the PLC using this basis [14].

The purpose of this paper is to introduce a class of
exactly solvable models for CT.
In these models, the PLC (formed in a hard process)
is treated as a coherent superposition of
a finite number of hadronic states.
We then assert that, at the point of formation,
the PLC-baryon interaction vanishes, Eq. (7).
Expressing the matrix elements of this vanishing interaction
in a hadron basis and using
completeness leads to a new set of sum
rules yielding relationships between elastic and inelastic soft hadronic
amplitudes. These relationships specify the form of the baryon-nucleon
scattering matrix, so that one avoids  assuming
a specific interaction  at the quark level.
Such an idea has been
applied first to CT in photoproduction of charmed
particles [7]. See also Refs.~[9,16].

 These models yield
new relations
between soft and hard processes.
Since a sufficiently hard amplitude can be calculated in pQCD,
the present work may represent the beginning of a new method to
calculate soft amplitudes: the generalization of dispersion sum rule
approaches to the physics of scattering processes.

Our models involve the full transition matrix ${-i\over 2E}\widehat T$
for $XN\to X'N$ reactions,
where $X,X'$ represents any baryonic state. (To understand the factors,
see below Eq. (11).) A very large number of
unmeasured matrix elements are then parameterized. We avoid an
unnecessarily large number of parameters by imposing constraints based on
existing data. In particular, using the recent analysis [17] of
matrix elements of $\widehat T^n$ in
high energy
proton-proton diffractive processes leads to significant constraints.
This makes the models more realistic than otherwise.

Here is an outline of the rest of the text. Some necessary formalism is
recalled in Sect.~2. The new sum rules and models are introduced in
Sect.~3. The models are constrained by diffractive dissociation data in
Sect.~4. Results are presented in Sect.~5. A few summary remarks form
Sect.~6.

\vfill \eject
\noindent 2. \ut{Scattering Formalism}

We begin the technical discussion by recalling  some
scattering formalism for relevant reactions.
To be specific, we discuss the (e,e'p) reaction for quasielastic
kinematics [4]. Other reactions could easily be considered using only modest
extensions of our equations. The (e,e'p) reaction proceeds when a
proton $|N\rangle$, bound in a shell model orbital $\alpha$,  absorbs a virtual
photon.
The absorption operator is the hard scattering operator $T_H(Q^2)$, and
a PLC is formed:
$$|PLC\rangle = T_H(Q^2)\;|N\rangle. \eqno(1)$$
The $PLC$ is thus a superposition of (a generally infinite number of)
hadron  states having baryon number of a nucleon. In practice, we
restrict  ourselves by considering the contribution of either one or
two resonances (or excited states) of the nucleon.
Such an approximation is
an effective method to treat convergent
dispersion integrals (see e.g. Ref. [18]) and greatly simplifies the
resulting models.

The amplitude for the proton knockout process is given by
$${\cal M}_\alpha = \langle N,\alpha|\;T_H(Q^2)\;|{\bf\Psi}_{N,\vec p}\rangle,
\eqno(2a)$$
where $|{\bf\Psi}_{N,\vec p}\rangle$ is the scattering wave function. The
subscripts $N$ and $\vec p$ Refer to the boundary condition that the
detected baryon is a proton of momentum $\vec p$.
For quasielastic kinematics, $\vec p$ is essentially parallel to
the direction of the virtual photon.  The measured cross section is
given by
$$\sigma\sim\sum_{\alpha,occ}\;|{\cal M}_\alpha|^2,\eqno(2b)$$
where $\sigma$ is obtained by integrating the angular distribution of
the outgoing  proton.  The sum is over all occupied shells.
We shall be concerned with ratios of $\sigma$ to the corresponding
quantity $\sigma^B$ which is evaluated by using a plane wave function,
instead of $|{\bf\Psi}_{N,\vec p}\rangle$, in Eq. (2a).

The coordinate space representation of $|{\bf\Psi}_{N,\vec p}\rangle$ is
a column vector ${\bf\Psi}_{N,\vec p}(\vec R)$ where $\vec R$ is the
distance between the center of mass of the baryon and the nuclear center.
The component of $\vec R$ parallel to $\vec p$ is denoted $\vec Z$.
Then $\vec R=\vec B +\vec Z$ and $\vec B $ is the impact parameter.
The different entries in the column vector Refer to the different baryonic
components or
channels ($N^*$, $\Delta$ etc.) of the scattering wave function.
A baryonic basis is used, so
we introduce an operator, $\widehat V$, to  describe the
baryon-nuclear interaction. The operator $\widehat V$ is obtained by
summing the baryon-nucleon transition matrix (describing the production of
baryon resonances) $\widehat T$ over the nucleons
and taking the
nuclear expectation value of the result.
The effects of correlations between the nucleons in the nucleus are
not important in the present context and are ignored.
Then $\widehat V$ is proportional to
the nuclear density, $\rho$, and $\widehat V={-i\over 2E}\widehat T \rho$.

We use a generalized eikonal approximation
to deduce the equation describing PLC propagation through the
nucleus. Spin-dependent effects are ignored here, so we start with
the relativistic Schr\"oedinger
equation and optical approximation:
$$(\sqrt{-\nabla^2 +\widehat M^2} + \widehat V)|{\bf\Psi}_{N,\vec p}\rangle =
E|{\bf\Psi}_{N,\vec p}\rangle,\eqno(3a)$$
where
$\widehat M^2$ is the baryon mass operator squared, $E=\sqrt{p^2 +M^2}$
 and $M$ is the nucleon
mass.
As noted above, the potential $\widehat V$ is calculated using amplitudes for
$NN$ interactions leading to production of resonances, cf.~Eq.(8).
We will consider situations in which $p>1 GeV/c$.
Then Eq.~(3a) may be simplified by multiplying both sides by the
operator $\sqrt{-\nabla^2+\widehat M^2}$ and using  Eq.~(3a)
to eliminate terms proportional to the square root operator. This gives
$$(-\nabla^2 + \widehat M^2)|{\bf\Psi}_{N,\vec p}\rangle=
(p^2 +M^2 - 2E\widehat V)|{\bf\Psi}_{N,\vec p}\rangle,
\eqno(3b)$$
in which terms proportional to $(\widehat V)^2$
and a gradient of $\widehat V$ are ignored.
This is valid at high energies $E$ and is consistent with the eikonal
approximation.

The eikonal approximation to Eq. (3) is developed by defining a new column
vector~$\bf\Phi$
$${\bf\Psi}_{N,\vec p}(\vec R) =e^{i\widehat pZ}{\bf\Phi}(\vec R),\eqno(4)$$
where
the operator $\widehat p$ acts in the
space of baryons:
$\widehat p^2 = p^2+M^2-\widehat M^2$.
The notation has been simplified by ignoring the subscripts $N,\vec p$
for $\bf\Phi$. The use of Eq. (4) in Eq. (3b) leads to
$$2i \;{\partial{\bf\Phi}(\vec R)\over\partial Z} =
 \widehat U\;{\bf\Phi}(\vec R)
,\eqno(5a)$$
where
$$\widehat U \equiv e^{-i\widehat p Z} \; {{2E}\over{\phat}} \widehat V
\;e^{i\widehat p Z}.\eqno(5b)$$
The basic tools to compute a wide variety of nuclear reactions,
suggested  in [9],  are
illustrated in Eqs. (1-5). Some applications and a more detailed discussion
of the formalism are
given in Ref. [19].
\vfill \eject
\noindent 3.\ut{Sum Rules and Models}

Now we turn to the new models.
The simplest is defined by
describing
the $PLC$
as a superposition of two states
$$|PLC \rangle = \alpha |N\rangle + \beta |N^* \rangle. \eqno (6)$$
In general, $\alpha$ and $\beta$ are complex functions of $Q^2$
representing elastic  and inelastic form factors.
We shall see that only ratios enter in our evaluations,
so we treat these functions as constants
 (In Ref.~[14] , $\alpha=\beta$).
Eq. (6) represents the small-sized wavepacket at the instant of
formation. As the wave-packet moves through the nucleus
each component acquires a different phase.
Thus the   size and other properties change.
In our formalism, the effects of such phases appear
in the exponential factors of Eq. (5b).

We determine the interaction $\widehat U $
by demanding  that
$$\widehat U (\vec B, Z = 0) |PLC \rangle = 0. \eqno (7a)$$
This relationship is equivalent to a set of sum rules. To see this,
use completeness to express the $PLC$ in terms of a
set of hadronic states $\mid m \rangle$, and take the overlap with one of
those, $\mid n \rangle$. This gives
$$\sum_m\langle n \mid \widehat U(\vec B,Z=0) \mid m \rangle
\langle m \mid PLC \rangle=0.\eqno (7b)$$
Thus one obtains a different sum rule for each state $\mid n \rangle$.
These equations can be developed in different ways. One can take
$\mid n\rangle$ to represent the nucleon and then break up the sum into
the discrete nucleon term  and a continuum for the other states[16].
Here we represent $\widehat U(\vec R)$ as a two- or three- dimensional
basis.

We start with the  two-state basis in which the $PLC$ is given by Eq.~(6).
Then the sum rules of Eq.~(7) are given by
an explicit expression:
$$\widehat U (\vec B, Z) = - i \sigma \rho(\vec R) \pmatrix {
1 & - {\alpha \over \beta}  e^{i \Delta p Z} \cr
-{\alpha^* \over \beta^*} e^{- i \Delta p Z} & \mid {\alpha
\over \beta}\mid^2 \cr }. \eqno(8)$$
\noindent Here, $\Delta p= p_{N^*}-p$, $p_{N^*}=\sqrt{p^2+M_N^2-M_{N^*
}^2}$.  The real part of the nucleon-nucleon
scattering amplitude is neglected here, as is valid
at the high energies ($p> 1 GeV/c$) we
consider. This is understood in terms of Regge theory:
keeping vacuum pole exchanges, and neglecting secondary Regge poles is a
good approximation.
Thus the nucleon-nucleon scattering amplitude is
proportional to the total cross section, $\sigma\approx$ 40 mb, so
that one element
of the $\widehat U$ matrix is determined by  data. In addition,
time-reversal  invariance is used to determine the
relationship  between the off-diagonal matrix elements.

The resonance mass $M_{N^*}$ and the related value of $\Delta p$ controls
the energy or momentum $p$  for which color transparency occurs. If the
value of $M_{N^*}$ is small enough so that $\Delta p Z$ can be
 ignored, the interaction $\widehat U$ vanishes and color transparency
occurs. We do not study the appropriate values of $M_{N^*}$ here; the work
of Ref.~[16] argues that reasonable values of $M_{N^*}$ are not very large
$\approx 1.7 GeV$.

The interactions of $\widehat U$ of Eq. (8) vanish at Z=0, increase with Z
to a maximum at $\Delta pZ = \pi$, and vanish again when
$\Delta pZ = 2\pi$. Thus the size of the PLC changes as it
moves through the nucleus.

We stress that, in this model,
the ratios of the soft amplitudes for $N^*$ to $N$ production in
nucleon-nucleon scattering are equal to ratios of hard form factors.
This is an example of a non-trivial
relationship between soft and hard processes that one expects to
exist in $QCD$.  In general, $\widehat U$ should act in a large
space and these many
states could complicate a more realistic treatment.  It may be
easier to examine the $c \bar c$ system (J/$\Psi,\Psi')$
[7] to discover these soft-hard
relationships.

The reader may be puzzled by the relations between the soft matrix
elements of the optical potential operator $\widehat U$ and the
hard baryonic (transition) form factor matrix elements,
$\alpha, \beta$. It is therefore worthwhile to exemplify  how
a
simple quark idea  can lead to unexpected relationships between
hadronic matrix elements. Consider
Bjorken scaling in deep inelastic scattering (DIS).
As explained in many reviews,
e.g. Refs.~[7,9], at low $x$ the DIS mechanism is
that of virtual photon ($\gamma^*$)
 decay into a $q \bar q$ color singlet pair which interacts with the target
nucleon. If the quark and anti-quark are closely separated,
gluon emission   effects are approximately cancelled
(color screening CS). This CS  is the essential element needed
to obtain scaling at low $x$ and is also necessary for
CT to occur.
Color screening is most simply expressed in terms
of quarks and gluons, but a
 hadronic basis may also be used.
The cross section for
$\gamma^*$-nucleon scattering is given by :
$$\sigma(\gamma^*N)=\sum_{n,r} {c_n \langle n\mid  T^{SOFT}\mid r\rangle
c_r^*\over (Q^2 + m_n^2)(Q^2+m_r^2)},\eqno(9)$$
when expressed in terms of hadronic matrix elements.
This equation describes the process of
$\gamma^*$ transition to a hadronic state n (of matrix element $c_n$);
interaction
with the target nucleon converts the state $n$ to a state $r$ ($T^{SOFT}$);
followed by conversion to the $\gamma^*$ ($c_r^*$).

The sum of the diagonal transitions is related to the $\gamma^*$
polarization operator as
$$\Pi(Q^2)=\sum_{n}{|c_n|^2\over(m_n^2 + Q^2)},\eqno(10)$$
which according to a QCD quark-loop calculation behaves as
$\sim Q^2 lnQ^2$.
The Bjorken scaling, which is well satisfied experimentally, requires
$\sigma(\gamma^*N)\sim {1\over Q^2}$.
At the same time, comparing the expressions for $\sigma(\gamma^*N)$ and
$\Pi(Q^2)$ shows that keeping only
the diagonal contribution leads to
$\sigma(\gamma^*N)\sim {1\over Q^2} \Pi(Q^2)\quad\sim lnQ^2$ - the
so-called Bjorken paradox.
The quark
ideas and experimental results are that this $ln Q^2$ term must vanish.
This occurs
only if
$$\sum_{n,r}c_n \langle n\mid  T^{SOFT}\mid r\rangle
c_r^*=0.\eqno(11)$$
This result as well as, for example, that of Eqs. (7) and (13)
below would be very difficult to guess at using only hadronic ideas.
(QCD logarithms are ignored for simplicity here.)
The above relation  is an
example, other than Eq.~(7) of an unexpected relationship between hadronic
matrix elements.

Thus perhaps Eq.~(7) is not so surprising; and we shall proceed by
evaluating its consequences.
Before doing so, it is necessary to to be aware that
two-state models are not realistic.
 One cannot model a wave packet of
very small size as a coherent sum of only two eigenfunctions corresponding
to normal-sized systems. For example, one can make $\langle r^2\rangle$
 vanish for a superposition of ground and first excited states
bound in a given potential. But then $\mid\langle r^4\rangle\mid$
would be rather large.
Moreover, diffractive dissociation data [20] for the pp$\to$pX reaction show
that many states $X$ (including a resonance region
and a higher mass continuum) are excited.

 We therefore make a
three state model.  ${Each}$ of
the proton, the resonance region and high mass region is represented
as one state in this model.
The derivation of the three-state model is as before.
This $PLC$ is defined as
$$\mid PLC_3 \rangle = {\alpha \mid N \rangle} +{\beta \mid N^* \rangle}
+ {\gamma
\mid N^{**} \rangle}.\eqno(12)$$
The soft operator $\U_3$ is again defined so that
$\U_3(\vec B,Z=0) \mid PLC_3 \rangle =0$.  This implies that
$$\U_3(\vec B,Z=0) = -i\sigma\rho\left(\matrix{1&-{\alpha\over\beta}+{\gamma
\over\beta}  \epsilon&-\epsilon\cr
-{{\alpha^*}\over{\beta^*}}+{{\gamma^*}\over{\beta^*}}\epsilon^*&\mu&{
{\mid\alpha\mid^2}\over{\beta^*\gamma}}-{{\alpha\gamma^*}\over{\beta^*
\gamma}}\epsilon^*-\mu {\beta\over\gamma}\cr
-\epsilon^*&{ {\mid\alpha\mid^2}\over{\beta\gamma^*}}-{{\alpha^*\gamma}\over
{\beta\gamma^*}}\epsilon-\mu{ {\beta^*}\over{\gamma^*
}}& \mu {{\mid\beta\mid^2}
\over{\mid\gamma\mid^2}}-{ {\mid\alpha\mid^2}\over{\mid\gamma\mid^2}  } + 2
\Re ( {\alpha\over\gamma}\epsilon^*)
\cr}\right).\eqno(13)$$

In this three state model, the free parameters are taken to be the masses of
the two excited states, $M_{N^*}$ and $M_{N^{**}}$, the excited diagonal
``optical factor" $\mu$, the ratios $\alpha\over\gamma$, $\beta\over
\gamma$ and, the coupling factor $\epsilon$.
The quantities $\mu$, $\nu \equiv  \mu {{\mid\beta\mid^2}
\over{\mid\gamma\mid^2}}-{ {\mid\alpha\mid^2}\over{\mid\gamma\mid^2}  } + 2
\Re ( {\alpha\over\gamma}\epsilon^*)$ are constrained to be greater than zero
by unitarity.
Further, at high energies it is well known that scattering amplitudes are
predominantly imaginary.  therefore, we choose our parameters to be purely
real so that ${\widehat U}_3$ is purely imaginary.

The results of this model depend upon
six (${\alpha\over \gamma}$, $\mid {\beta\over\gamma}\mid$,
$\epsilon$,  $\mu$,
 $M_{N^*}$, $M_{N^{**}}$) parameters.
The reader may wonder if this is too many.
We note that the matrix
of Eq.~(13) is a model of the entire matrix of baryon-baryon scattering
amplitudes. As such it represents an infinite number of
matrix elements. Our point is that Eq.~(7) provides a powerful and
testable constraint on this matrix.
\vfill \eject
\noindent 4. \ut{Contraints on model parameters}

One may improve the model by imposing
constraints which relate the
parameters to soft process observables.
The matrix of Eq.~(13) is related to
ratios of inelastic to elastic proton-nucleon cross sections. Making
a detailed comparison between the parameters used here and data seems
unrealistic. Instead we employ information regarding fluctuations of
hadronic cross sections obtained recently by Bl$\ddot{\rm a}$ttel et al.~[17].
Those authors used the Good-Walker [21] cross section eigenstate
formalism to analyze inelastic shadowing and diffractive dissociation
data in terms of moments of the transition matrix,
$\langle N\mid \widehat T^n\mid N \rangle $ where
$$\widehat T \equiv e^{i\phat Z} {{i{\widehat U}}\over{\rho}} e^{-i \phat
Z}.\eqno(14)$$
With a purely imaginary $\widehat U$,
$\widehat T$ represents the imaginary part of the baryon-nucleon transition
matrix; as such it is independent of $\vec B$ and $\vec R$.
The results of Bl$\ddot{\rm a}$ttel et al. (for 200 GeV protons) are that
$$\langle N\mid \widehat T^2\mid N\rangle=(1+\omega_{\sigma})\sigma^2,
\eqno (15a)$$
with $\omega_{\sigma}=0.2$. Furthermore
$$\langle N\mid \widehat T^3\mid N\rangle-\sigma \langle N\mid \widehat T^2
\mid N\rangle=\kappa_{\sigma}\sigma^3,\eqno(15b)$$
where $\kappa_{\sigma}\sim 2 \omega_{\sigma}$.

At lower energies, the relevant number of (parton) degrees of freedom is
less than that at higher energies. Thus a smaller value of
$\omega_{\sigma}$ is expected. Here we take
$$\omega_{\sigma}\lsim 0.1,\qquad
\kappa=0.2.\eqno(16)$$
These equations strongly constrain
 the matrices of Eqs. (8) and (13). Indeed, it only
takes
a moment to rule out all two-state models with purely
imaginary amplitudes.
For the two-state model of Eq. (8), the evaluations of Eq. (15) yield
$$1 +\mid \alpha/\beta\mid^2= 1+\omega_{\sigma}\eqno(17a)$$
and
$$(1 +\mid \alpha/\beta\mid^2)^2-(1 + \mid \alpha/\beta\mid^2)=\kappa_{\sigma}
\approx 2\omega_{\sigma}.\eqno(17b)$$
The use of (17a) in (17b) then leads to the
result $\omega_{\sigma}=1$.
This failure of two-state models is consistent with the intuition that
the
expectation value of $r^4$ will be large even if that of $r^2$ vanishes.
All of the results shown below satisfy the constraints of Eqs. (15) and (16).

It is also reasonable to see if
the quantities $\beta/\alpha$ and $\gamma/\alpha$
represent ratios of excited state to proton form factors. In principle,
these ratios are measurable. In the present model,
$\mid N^{*,**}\rangle$ represents a coherent superposition of resonances,
so that it is not possible to compare $\beta/\alpha$ and $\gamma/\alpha$
measured baryon resonance form factors. Below,
we use values of $\beta/\alpha$ of the order of three or five. The
measured ratios for single resonances are of the order of unity [22], so that
our state $\beta \mid N^*\rangle$ is consistent with a superposition of a
few resonances.
\vfill \eject
\noindent 5. \ut{Results}

We now turn to the evaluation of the model.
Analytic solutions may be obtained
by assuming that the $PLC$
is produced at $Z$ = 0 and propagates a distance $L$ through uniform
nuclear matter of
constant density,
 $\rho=\rho_0$ = 0.166 $fm^{-3}.$
For many sets of parameters of this three-state model, Eq. (5) reduces to
three coupled first order differential
equations, convertible
into a third order equation with constant coefficients.  The
solution is an exponential. The eigen-`energies' can be found exactly,
because a general cubic equation is soluble.  The solution is useful
for checking the numerics, but is not illuminating
and omitted.

One may mimic the physics  of the (e,e'p) reaction with
this one-dimensional model by letting
the distance $L$ be the average
distance across a nucleus, ${L= 0.6 4 \, A^{1/3} \,\,fm}$, and by
setting $\vec p$ equal to the momentum of the virtual photon.
 The equations which result from taking a constant nuclear density are very
similar to the well
known molecular beam physics equations which lead to Rabi's
formula [23].
We compute the
probability that, after traversing a distance $L$ through the nuclear matter,
a nucleon (or an excitation) will be detected.
Then the observables are the probabilities,
${\cal P}_N(L) = {\mid \langle N | \Phi (L) \rangle \mid ^2}$.
Similarly, the probability to produce an $N^*$ or $N^{**}$ is given by
 ${\cal P}_{N^{*(**)}}(L) = {\mid \langle N^{*(**)}| \Phi(L) \rangle \mid ^2}$.

The results of the nuclear slab calculations are shown in Fig.~1.
It is necessary to discuss how the parameter sets were
chosen.
We choose $M_{N^*}=1.4$ GeV and $M_{N^{**}}=2.4 $ GeV
as representative of the resonance and continuum regions. We make no
attempt here
to fine-tune or constrain these mass values. It is clear that using
larger values will lead to larger effects of CT.
After the masses were chosen, $\alpha/\gamma$ was chosen to be either
large (3), moderate (1),
or small (1/3). Then the values of $\beta/\gamma$ and $\epsilon$ are
chosen over a broad grid of values. The values of $\mu$ and $\nu$
(third diagonal matrix element of Eq.~(13))
are then uniquely determined by the constraints of of Eq. (16).
Finally, only those
parameters sets leading to values of $\mu$ of the order of unity are kept.
That term represents the ammount of absorption in the $N^*$-nucleon
diagonal scattering, which we take
to be of the same order as that of the proton-nucleon.

As shown in Fig.~1, the overall size of the probabilities,
 e.g. ${\cal P}_N(L)$,
depends mainly on the importance
of the second excited state (value of $\gamma$).
Results with large, moderate, and small values of $\alpha/\gamma$ are
shown in Fig.~1.
All of the parameter
sets discussed above lead to significant sizes of
the probabilities.
We call attention to the interesting oscillations
(wiggles) as a function of the photon three-momentum,
Figs.~1c and~d.
Furthermore, effects of constructive
interference can lead to cross section ratios (transparencies)
larger than unity (Fig.~1d), which may occur if $\epsilon$ is negative.

 The next step is to generalize this
one-dimensional calculation to that of the
three-dimensional formalism of Eqs. (1-5).
This is done by allowing the nuclear density $\rho$ to be
a function of $\vec B$ and Z.
Then, within the eikonal approximation, the column vector
${\bf\Phi}$ ($\mid {\bf\Psi}_{N,\vec p}\rangle$) is known for that $\vec B$.
The cross sections $\sigma$ and  $\sigma^B$  are obtained by using
the numerical solution for
${\bf \Psi}_{N,\vec p}$ (to compute $\sigma)$
 or plane wave  functions (to compute $\sigma^B$)
in the matrix elements of Eq.~(2).
The density
$\rho(\vec R)$ is described as a standard Wood-Saxon form.
We also allow the masses of the
baryon excitations to have a small imaginary part, 75 $MeV$.
(This is not significant numerically, but simulates a typical
resonance width of $\sim 150\; MeV$.) For details regarding the
bound state wave functions  and other aspects, see Ref.~14.

The  use of the parameters of
Fig.~1 lead to the very different ``three-dimensional" results for the
ratios $\sigma/\sigma^B$ shown in Fig.~2.
 The wiggles are smoothed out,
since the necessary average over impact parameters causes the
path length to take on a
continuous series of values.

The loss of the oscillations shows that averaging over impact parameters
makes the expansion of the PLC more classical in the sense
that effects of quantum mechanical interference  are suppressed.
One may search for the wiggles by considering processes
in which the amplitude depends on higher powers of $\rho$.
Then small impact parameters would be emphasized and the important values
of  the path
length would be less spread out.
One example, could be the  production of backward particles in quasielastic
processes. Another could be (p,2p) sub-processes in heavy ion collisions. This
notion will be pursued in other work.

There are other noteworthy features in the comparison of Figs. 1 and 2.
The ratios $\sigma/\sigma^B$ are generally lower than the corresponding
probabilities. This is caused by the diffuse nature of the nuclear surface.
Even so, the ratios $\sigma/\sigma^B$ are much larger than the standard
Glauber calculation (keeping only the nucleon channel
in the scattering wave function) labelled by DWBA in the Figures.

Another  point of interest is that
 in Figs. 1a,b the apparent transparency of the second resonance is
quite high, but this does not occur in Figs. 2a,b. For the parameter sets
of those figures,
$\epsilon$ is very small. Thus the production of the $N^{**}$ is dominated
by feeding from the $N^{*}$. This $N^{**}$ production
can only occur near the edge of the slab
or (Fig.~1) or near the edge of the nucleus (Fig.2b) because
the
value of the second excited state diagonal element is very large.  Indeed,
in Fig. 1a~(2a), $\nu = 130$ and in Fig. 1b~(2b),
$\nu=14.2$.  In contrast with the full nuclear matter density
of the slab,
the
density and bound state wavefunctions are both quite small on the surface
of the three-dimensional nuclear target. Thus the edge effects of
$N^{**}$ production through the $N^{*}$ channel occur only  for the nuclear
slab.

The  final result we show
concerns using
a set of parameters closely related to the model of
Ref. [14]; see Fig.~3.
  As above, the solid curves denote the cross section ratio for a nucleon.
The dashed
and dotted curves represent the ratios for the first and second
excited states in this three state model.
The exponential approximations (EA and EA$^*$) of Ref. [14]
sum many higher order terms by exponentiating the first order (in $\widehat U$)
result.
The results here verify the accuracy of the exponential approximation, at
least for the parameters of Fig.3.  More than that, this result shows that
a three dimensional baryon space is ``large enough'' to act as an infinite
dimensional one for the model of Ref. [14].  Further details are
discussed
in Ref. [19].

Many sets of parameters lead to results qualitatively similar to those of
Figs.~1-3. However, the use of a large value for the mass $M_{N^{**}}$
can suppress color transparency or push its appearance to a very high
energy.
\vfill \eject
\noindent 6. \ut{Summary}

A new set of models of
baryon-nucleon interactions is obtained from the assumption that a PLC
does not interact with baryons.
This involves new sum rules which relate hard and soft processes.
The PLC's motion through the
nucleus is governed by a solvable scattering wave equation.
The model is made more realistic by requiring consistency with certain
diffractive proton-proton scattering
observables, Eqs.(12,13). These constraints allow significant effects of
CT to occur. Furthermore,
interesting effects such as rapid oscillations with energy  and
transparencies greater than unity are also allowed.
Results are presented for the (e,e'p) and (e,e'N$^*$)
reactions, but the model can
be applied [9] to any study of color
transparency such as the (p,pp) reaction; the generalized
optical potentials can be used in any process that requires
nuclear scattering wave functions. For example, neutron-nucleus
total cross sections can be computed.
Another  possible further extension could be
to develop a new sum rule approach to learn about the
properties of the PLC. One may evaluate
matrix elements of $\widehat T^n$
in both
the quark-gluon and hadronic  bases. Since the two approaches must give
the same result, powerful constraints can be obtained.
Perturbative QCD or quark models can be used to get $\widehat T^n$ in the
quark-gluon basis and some hadronic results can be obtained from data.
Such new sum rules will be discussed elsewhere.
Thus, the present work could  lead to a new method to relate
the results of pQCD to soft processes.

We thank the DOE and NSF for partial support.
\vfill\eject
\noindent {\bf
References and Footnotes}
\item{1.} A.H.~Mueller, ``Proceedings of Seventeenth Rencontre de
    Moriond", Moriond, 1982 ed. J Tran Thanh Van (Editions Frontiers,
Gif-sur-Yvette, France, 1982)p13.
\item{2.} S.J.~Brodsky in Proceedings of the Thirteenth
intl Symposium
on Multiparticle Dynamics, ed. W.~Kittel, W.~Metzger and A.~Stergiou (World
Scientific, Singapore 1982,) p963.
\item {3.}A.S. Carroll et al Phys Rev Lett 61,1698 (1988; S. Heppelmann,
p. 199 in ``Nuclear physics on the Light Cone", ed. M.B. Johnson and
L.S. Kisslinger , World Scientific (Singapore, 1989).
\item {4.}SLAC Expt. NE-18, R. Milner, Spokesman.
\item {5.} A.S. Carroll {\it et al.}, BNL expt. 850.
\item{6.} F.E. Low, Phys. Rev. D12, 163 (1975);
 S. Nussinov Phys. Rev.  Lett 34, 1286 (1975);
J. Gunion and D. Soper Phys Rev D15,2617 (1977).
\item{7.} L. Frankfurt and M. Strikman, Progress in Particle and Nuclear
Physics, 27,135(1991);
L. Frankurt and M. Strikman, Phys. Rep. 160, 235 (1988).
\item{8.}  G.R.~Farrar, H.~Liu, L.L.~Frankfurt \& M.I.~Strikman, Phys.
Rev. Lett. 61 (1988) 686.
\item{9.} L Frankfurt, G.A. Miller \& M. Strikman, ``Color Transparency and
Nuclear Phenomena", to be published Comm. Nuc. Part. Phys. Sept. 1992.
1992 UWA preprint-40427-26-N91.
\item{10.} B.Z. Kopeliovich, Sov.J. Part. Nucl. 21, 117 (1990).
\item{11.} H-n Li and G. Sterman,``The perturbative Pion Form Factor with
Sudakov Suppression", 1992 preprint ITP-SB-92-10.
\item{12.}L. Frankfurt, G.A. Miller \& M. Strikman, ``Precocious dominance
of Point-like
configurations in Hadronic Form Factors", submitted to Nucl.
Phys. A. U.Wa. preprint 40427-16-N92.
\item{13.}A.H.Mueller Phys.Rep.73,237 (1981)

\item{14.} B.K.~Jennings and G.A.~Miller, Phys. Lett. B236, 209 (1990);
 B.K.~Jennings and G.A.~Miller, Phys. Rev. D 44, 692 (1991);
 G.A. Miller and B.K. Jennings p. 25 in "Perspectives in Nuclear Physics
at Intermediate Energies" Ed. S. Boffi, C. Ciofi degli Atti, M. Giannini,
1992, World Sci. Press Singapore;
G.A.  Miller, ``Introduction to Color Transparency",
in ``Nucleon resonances and Nucleon Structure",  G.A.
Miller, editor. To be published by World Sci., Singapore (1992);
B. K. Jennings and G.A. Miller,
Phys. Lett. B 274,442 (1992).
\item{15.}  B.Z.~Kopeliovich and B.G.~Zakharov, Phys. Lett. B264 (1991) 434.
\item{16.} B.K.~Jennings and G.A.~Miller, 1992 TRIUMF, U.Wa.  preprint
40427-24-N92,
``Realistic Hadronic Matrix Element Approach to Color Transparency"
submitted to Phys. Rev. Lett.
\item {17.} B. Bl$\ddot{\rm a}$ttel , G. Baym ,L.L. Frankfurt, H. Heiselberg
and M.~Strikman, ``Hadronic cross-section fluctuations", Univ. Illinois 1992
preprint, P92-6-66.
\item{18.}
V. de~Alfaro, S.~Fubini,G.~Furlan,and C.~Rosetti ``Currents in Hadrons"
North-Holland (Amsterdam) 1973.
\item{19.} W.R.~Greenberg and G.A.~Miller
``Multiple-Scattering Series for Color
Transparency' UWA preprint 1992, 40427-23-N92, submitted to Phys Rev. D.
\item{20.} K. Goulianos, Phys. Rep. 101, 169 (1983)

\item{21.} M.L. Good and W.D. Walker, Phys. Rev. 120,1857 (1960).
\item{22.} P. Stoler, Phys. Rev. Lett. 66,1003 (1991); Phys. Rev. D44,73
(1991).
\item{23.} Rabi's formula is discussed in Chapt. 5 of ``Modern
Quantum Mechanics" by J.J. Sakurai, Addison Wesley, Redwood City, CA 1985.

\vfill \eject
\noindent {\bf Figure Captions}
\bigskip
\item{1.}Nuclear slab probabilities, $^{12}C(e,e'p)$. The
Solid- nucleon production. Dashed- $N^*$ production. Dotted-$N^{**}$
production.  The parameters $(\mu,\nu)$ (uniquely determined by $\epsilon$,
$\alpha/\gamma$, and $\beta/\gamma$) for a,b,c, and d are
(1.6, 130), (1.6, 14.2), (1.5, 1.2),
(0.42, 0.96) respectively.
\medskip
\item{2.} Three dimensional shell model ratios of cross sections.
$^{12}C(e,e'p)$. Solid- nucleon production. Dashed- $N^*$ production.
Dotted-$N^{**}$
production. DWBA-proton cross section ignoring CT effects is also shown as
the straight solid line. The parameters
$\mu$ and $\nu$ are as in Fig. 1.
\medskip
\item{3.} Three dimensional shell model ratios of cross sections
$^{12}C(e,e'p)$. The curves are defined in the text.  See also Ref. [19].
\medskip

\bye